\documentclass{aa}  
\bibpunct{(}{)}{;}{a}{}{,}

\usepackage{afterpage,natbib,lipsum}
\usepackage{graphicx}
\usepackage{multirow}

\usepackage{txfonts} 
\begin{document}

   \title{Imprints of the large-scale structure on AGN formation and evolution}


   \author{Nat\`{a}lia Porqueres\inst{1,2} \and Jens Jasche\inst{1} \and Torsten A. En{\ss}lin\inst{2,3} \and Guilhem Lavaux\inst{4,5} }

   \institute{Excellence Cluster Universe, Technische Universität München, Boltzmannstrasse 2, 85748 Garching, Germany 
   		\and
   		Max-Planck-Insitut f\"{u}r Astrophysik (MPA),
             Karl-Schwarzschild-Strasse 1 , D-85741 Garching, Germany
         \and
         Ludwig-Maximilians-Universität München, Geschwister-Scholl-Platz 1, 80539, München, Germany 
         \and
         Institut Lagrange de Paris (ILP), Sorbonne Universit\'{e}s,
98bis boulevard Arago, F-75014 Paris, France
         \and
         Institut d’Astrophysique de Paris (IAP), UMR 7095, CNRS – UPMC Universit\'{e} Paris 6, Sorbonne Universit\'{e}s,
98bis boulevard Arago, F-75014 Paris, France             
         }

   \date{Received 20/10/2017; accepted 08/01/2018}

 
  \abstract
   {Black hole masses are found to correlate with several global properties of their host galaxies, suggesting that black holes and galaxies have an intertwined evolution and that active galactic nuclei (AGN) have a significant impact on galaxy evolution. Since the large-scale environment can also affect AGN, this work studies how their formation and properties depend on the environment. We have used a reconstructed three-dimensional high-resolution density field obtained from a Bayesian large-scale structure reconstruction method applied to the 2M++ galaxy sample. A web-type classification relying on the shear tensor is used to identify different structures on the cosmic web, defining voids, sheets, filaments, and clusters. We confirm that the environmental density affects the AGN formation and their properties. We found that the AGN abundance is equivalent to the galaxy abundance, indicating that active and inactive galaxies reside in similar dark matter halos. However, occurrence rates are different for each spectral type and accretion rate. These differences are consistent with the AGN evolutionary sequence suggested by previous authors, Seyferts and Transition objects transforming into LINERs (Low-Ionization Nuclear Emission Line Regions), the weaker counterpart of Seyferts. We conclud that AGN properties depend on the environmental density more than on the web-type. More powerful starbursts and younger stellar populations are found in high densities, where interactions and mergers are more likely. AGN hosts show smaller masses in clusters for Seyferts and Transition objects, which might be due to gas stripping. In voids, the AGN population is dominated by the most massive galaxy hosts.}
   


   \keywords{   large-scale structure of Universe --
                active --
                evolution --
                formation --
                Seyfert
               }
               
    \titlerunning{Imprints of the large-scale structure on AGN}
	\authorrunning{Porqueres et al.}

   \maketitle
%

\section{Introduction}
	Active galactic nuclei (AGN) are believed to have a significant impact on galaxy evolution. Black hole masses correlate with several properties of the host galaxy, including for example the stellar mass and the velocity dispersion. This suggests that black holes and their host galaxies have an intertwined evolution and that AGN affect galaxy evolution.
    	
	The large-scale environment also plays a role in galaxy evolution. It has been shown that star formation rates are influenced by the large-scale environment \citep[see e.g. ][]{Balo04, Eina05, Liet09, Chen13}. For example, colors and morphologies of galaxies depend on the density of their host cluster. \cite{Eina08} has shown that luminosity functions of elliptical galaxies are strongly affected by the environmental density amplitude. Since the AGN phase may be a short period in the evolution of all massive galaxies, an interesting question is how the large-scale environment affects AGN. Particularly, how the formation and properties of AGN depend on the environmental density.
    
    The standard model of AGN \citep{StandardModel} assumes that the energy they release is produced by the accretion of gas onto a central supermassive black hole. However, the coupling between black hole accretion  and star formation in AGN remains unclear. In order to determine how these two processes happen and whether they synchronize, some works focused on the study of AGN in individual voids since more violent processes such as gas stripping are less likely to occur at low densities. The earliest works \citep{Kirs81,Cruzen2002} were limited to individual voids and concluded that AGN properties are similar in voids and clusters. However, those studies could not exclude the possibility that the AGN observed in the studied void were not residing in filaments, which would mask the local environment in the void. 
    
    Statistically significant studies have emerged recently with the release of large surveys, in particular, the Sloan Digital Sky Survey (SDSS)  \citep{DR7}. It was shown that the occurrence of the AGN activity as a function of the environment depends on the properties of the host galaxy \citep{Kauf03c,Const08,Hwang11,Silver2008,Pimbb13}. \cite{Kauf03c} and \cite{Cons06} have reported that strongly accreting AGN are found predominantly in low density regions. According to \cite{Const08} void AGN in massive hosts exhibit stronger accretion and younger stellar emission than their cluster counterparts. \cite{Hwang11} found that the morphology of the host galaxy strongly affects the dependence of the AGN fraction with the environment. 
    
    The study of the AGN clustering signal, defined as the cross-correlation of AGN with reference galaxies, shows that active and inactive galaxies have the same environment on large scales \citep{Li2006,Jiang2016,Karh14} while they show an overdensity of nearby neighbors (distance <1 h$^{-1}$ Mpc). This indicates that halos hosting AGN and inactive galaxies have similar masses.
    
    \cite{Const08} have analyzed the environmental dependence of the properties of each AGN spectral type. Although some differences between spectral types can be explained by the orientation angles with respect to the line of sight since the disk obscuration affects the AGN spectra, \cite{Temp09} and \cite{Const08} suggested that the different spectral types may form a sequence in galaxy evolution. Since the dynamical evolution in low density regions is expected to be slower, this evolutionary sequence might appear as a dependence of the spectral type with the environmental matter clustering. The study of this dependence also suggests an evolutionary sequence from quasars to radio-loud galaxies \citep{Liet11}.
    
     The goal of this paper is to study the relation between AGN and their global environmental density field, particularly the dependence of the incidence and properties of different spectral types with the density. Unlike most of the previous works {\citep[e.g.][]{Cons06, Liet11,Karh14,Coziol17}, in which the density field is obtained from galaxy counts, we use a three-dimensional high-resolution density field obtained from a Bayesian reconstruction applied on the 2M++ sample \citep{Lava11,Jasc13,Lavaux2016}. 
     
    This paper is structured as follows: in Section 2 we present the method to reconstruct the large-scale density field. In Section 3 we detail the dataset and the spectral classification of AGN into Seyferts, LINERs, and Transition objects. In Section 4 we study the dependence of the properties and occurrence rates of the different types of AGN with the environment. We have investigated this dependence as a function of the density contrast as well as of a web-type classification based on the shear tensor, identifying cosmic web structures like voids, sheets, filaments, and clusters. We summarize our results in Section 5.

\section{Methodology} \label{sec:method}
This section provides a detailed overview of the methods 
used to study AGN properties in the cosmic large-scale structure (LSS).
We also provide a brief overview of the Bayesian inference method providing the three-dimensional density field used in this work.

\subsection{Bayesian large-scale structure inference} \label{sec:BORG}
This work builds upon previous results of applying the Bayesian Origin Reconstruction from Galaxies ({\bf{BORG}}) algorithm to the data of the 2M++ galaxy compilation \citep[see e.g.][]{Lava11,JASCHEBORG2012,JLW15,Lavaux2016}.  

The {\bf{BORG}} algorithm is a fully probabilistic inference method aiming at reconstructing matter fields from galaxy observations. This algorithm incorporates a physical model for gravitational structure formation, which allows inferring the three-dimensional density field and the corresponding initial conditions at an earlier epoch from present observations.

Specifically the algorithm explores a large-scale structure posterior distribution consisting of a Gaussian prior for the initial density field at an initial cosmic scale factor of $a=10^{-3}$ linked to a Poissonian model of galaxy formation at a scale factor $a=1$ via a second order Lagrangian perturbation theory \citep[for details see][]{JASCHEBORG2012}. The model accurately describes one-, two- and three-point functions and represents very well higher-order statistics, as it was calculated by \cite[see e.g. ][]{MOUTARDE1991,BUCHERT1994,BOUCHET1995,SCOCCIMARRO2000, 2013JCAP...11..048L}. Thus {\bf{BORG}} naturally accounts for the filamentary structure of the cosmic web typically associated with higher-order statistics induced by nonlinear gravitational structure formation processes. The posterior distribution also accounts for systematic and stochastic uncertainties, such as survey geometries, selection effects and noise typically encountered in cosmological surveys.
  
We also note, that the {\bf{BORG}} algorithm infers initial 3D density fields at their Lagrangian coordinates, while final density fields are recovered at corresponding final Eulerian coordinates. 
Therefore the algorithm accounts for the displacement of matter in the course of structure formation.

As mentioned above, in this work we have used inferred LSS properties previously obtained by applying the {\bf{BORG}}  algorithm to data of the 2M++ galaxy sample \citep{Lava11}. Three-dimensional matter density fields have been inferred on a cubic Cartesian grid of side length of 677.7 $h^{-1}$ Mpc consisting of  256$^3$ equidistant voxels. This results in a grid resolution of 2.6 Mpc h$^{-1}$. Further we assumed a standard $\Lambda$CDM model with the following set of cosmological parameters: $\Omega_m = 0.3175, \Omega_\Lambda = 0.6825, \Omega_b = 0.049, h=0.6711, \sigma_8= 0.8344 , n_s = 0.9624$ \citep{Planck2014}. We assume $H=100 h$ km s$^{-1}$ Mpc$^{-1}$. A slice of the final density field is shown in the upper panel in Fig. \ref{fig:field} with the AGN superimposed.

The observer is at the center of the box and for the sake of this work, we concentrate on a spherical region  with a radius of 120 Mpc h$^{-1}$. In Appendix \ref{ann:align} we describe the alignment of observed AGN coordinates with the density field. Since AGN are affected by the redshift space distortions, we do the analysis in the redshift space.

\begin{figure*}[t]
\centering
\includegraphics[width=\textwidth]{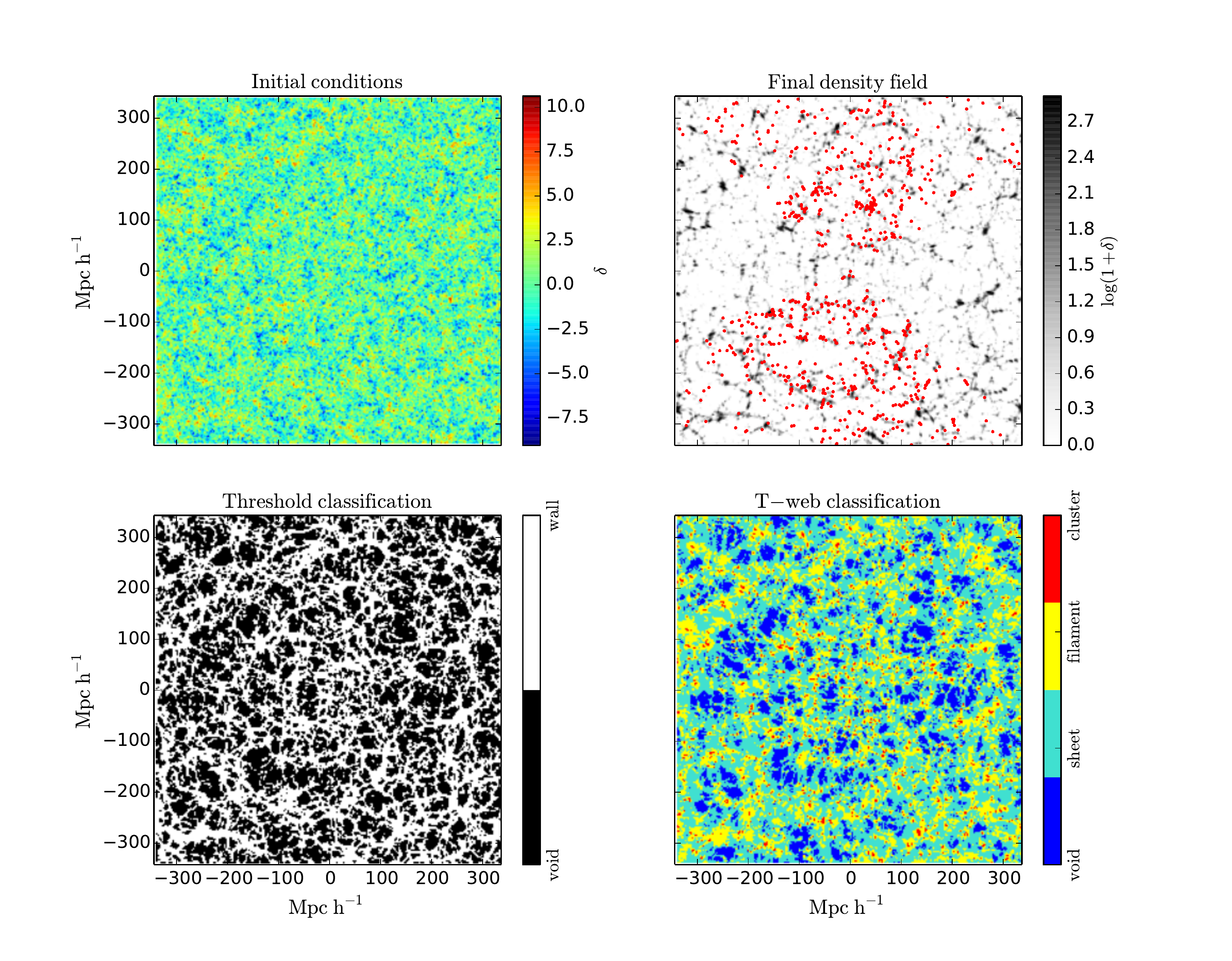}
\caption{Slice of the 3D density field. The upper panels show the initial and final density contrast. AGN are shown on top of the final density field. The bottom left panel shows the different structures defined by a threshold in the density contrast $\delta=-0.6$ following the approach in \cite{Const08}. However, \cite{Const08} defined the density field by smoothing the galaxy distribution which might result in thinner structures. The bottom right panel shows the cosmic web structures according to a web-type classifier based on the tidal shear tensor. \label{fig:field}} 
\end{figure*}

\subsection{T-web classification} \label{sec:Ts}
A detailed web-type classification can be achieved via a T-web analysis \citep[see e.g.][]{Hahn2007,ROMERO09,Lecq15}.
For a more complex classification of the large-scale environment we use a dynamic web classifier that dissects the entire large-scale structure into different structure types: voids, sheets, filaments, and clusters. This classification is based on the tidal shear tensor,
\begin{equation}
T_{ij} = \frac{\partial^2 \Phi}{\partial x_i \partial x_j},
\label{eq:ts_def}
\end{equation}
where $\Phi$ is the gravitational potential which can be obtained from the Poisson equation $\Delta \Phi (\vec{x}) = \delta (\vec{x})$. Different structures are classified according to the sign of the $T_{ij}$ eigenvalues. 

	The interpretation of this classification is straightforward because the sign of the eigenvalue defines whether a structure is expanding (negative) or contracting (positive) in the direction of the eigenvector. Therefore, the signature of the tidal tensor $T_{ij}$  defines the number of axes along which there is a plausible gravitational collapse or expansion. As summarized in Table \ref{tab:webtype}, when the three eigenvalues are negative, the region is defined as a void. If there is only one positive eigenvalue, it is a sheet (two-dimensional structure), while it is classified as a filament when there is only one negative eigenvalue. Clusters are identified as three positive eigenvalues. The result of the web-type classification is shown in the bottom right panel in Fig. \ref{fig:field}, in which the structures are defined at scales of 2.6 h$^{-1}$ Mpc. 
    
\begin{table}
\centering
\begin{tabular}{ c c c }
Web-type & Eigenvalues \\ \hline
Void & $\mu_1<0,\mu_2<0,\mu_3 < 0$ \\ 
Sheet & $\mu_1<0,\mu_2<0,\mu_3 > 0$ \\ 
Filament & $\mu_1<0,\mu_2>0,\mu_3 > 0$ \\ 
Cluster & $\mu_1>0,\mu_2>0,\mu_3 > 0$ \\ \hline
\end{tabular}
\caption{Web-type classification according to the eigenvalues $\mu_i$ of the shear tensor. \label{tab:webtype}}
\end{table}

\subsection{Computation of the abundance and occurrence rate} \label{ann:abundance}
	The abundance is defined as the number density of objects as a function of the environmental density. Since the mean density regions are more extended than voids and clusters, we needed a volume correction which is computed as the fraction of volume with a given density $v(\delta) = \frac{V(\delta)}{V_\mathrm{total}}$. Then the logharithm of the number density is computed as 
    \begin{equation}
	 \ln\left(\frac{N}{v}\right)=\ln\left(\frac{N_\text{objs}(\delta_\text{min}<\delta<\delta_\text{max})}{v(\delta_\text{min}<\delta<\delta_\text{max})}\right)
	\end{equation}
	where $\delta_\text{min}$ and  $\delta_\text{max}$ define the density bin. 
    
    The occurrence rate is defined as the number of objects of a spectral type in a given density divided by the total number of objects in this density, for example $N_\mathrm{Sy}(\delta)/N_\mathrm{AGN}(\delta)$. These quantities are computed with the Blackwell-Rao estimator (see Section \ref{app:BR}).

\subsection{Blackwell-Rao estimator} \label{app:BR} 

The Markovian samples described in Section \ref{sec:BORG} permits us to provide an uncertainty quantification of our results. We employed the Blackwell-Rao estimators 
	\begin{equation}
	\langle x|d \rangle = \frac{1}{N}\sum_i x_i,
	\end{equation}
in which $x$ is the quantity we want to study, for example the abundance, $N$ is the number of Markovian samples and $d$ the observations. This estimator was necessary because we were studying nonlinear functions of the density and $\langle f(\delta)\rangle \approx\langle f(\frac{1}{N}\sum_i\delta_i)\rangle$ only when the quantity of study $f$ is linear. It also allowed us to calculate the uncertainty as 
\begin{eqnarray}
\langle(x-\langle x \rangle)^2|d\rangle = \frac{1}{N}\sum_i (x_i^2+\sigma_i^2)-\langle x|d\rangle^2.
\end{eqnarray}

All the quantities in this study have been computed using the Blackwell-Rao estimators on 50 Markovian samples. 

\section{Data}
In this section, we describe the datasets employed in this work and provide a description of derived quantities and data selections.

	\subsection{AGN catalog}
   
   The analysis presented in this work is based on the catalog of the MPA/JHU collaboration \citep{Kauf03c}, which is a subset of the Sloan Digital Sky Survey Data Release Seven \citep{DR7}. This catalog contains properties of the host galaxy such as the stellar mass and the dust attenuation strength as well as the velocity dispersion, being proportional to the mass of the central black hole \citep{Ferr00}. It also contains the amplitude of the 4000 \AA{} break and the strength of the H$\delta$ absorption line which are indicators of the mean stellar age and recent starbursts. As a proxy of the morphology of the host galaxy, we will use the concentration index $C$ and the effective stellar surface mass-density. \cite{Strateva01} have shown that early-type galaxies have $C>2.6$ while spiral and irregular galaxies have $C<2.6$. The [OIII] line is produced by ionizing radiation that escapes along the polar axis of the dusty obscuring structure where it photo-ionizes and heats the medium \citep{Heck14}. Therefore the [OIII] luminosity is related to the nuclear activity and so it is an indicator of the accretion rate \citep{Kauf03c,Heck04,Heck14}. Following \cite{Const08}, we consider that AGN with $L_\mathrm{[OIII]}<10^{39}$ erg s$^{-1}$ are relatively weakly accreting objects.
   
   According to the standard model \citep{StandardModel}, active galaxies are classified as type 1 AGN when the broad emission-line region is observed directly while type 2 are those for which it is obscured by the interstellar medium. In type 1, the continuum is dominated by nonthermal emission and thus it is difficult to estimate their host galaxy properties. For this reason, type 1 objects are excluded from the MPA/JHU catalog \citep{Kauf03c}. 
          
	\subsection{AGN classification}    
    
\begin{figure}[t]
\centering
\includegraphics[width=3.7in]{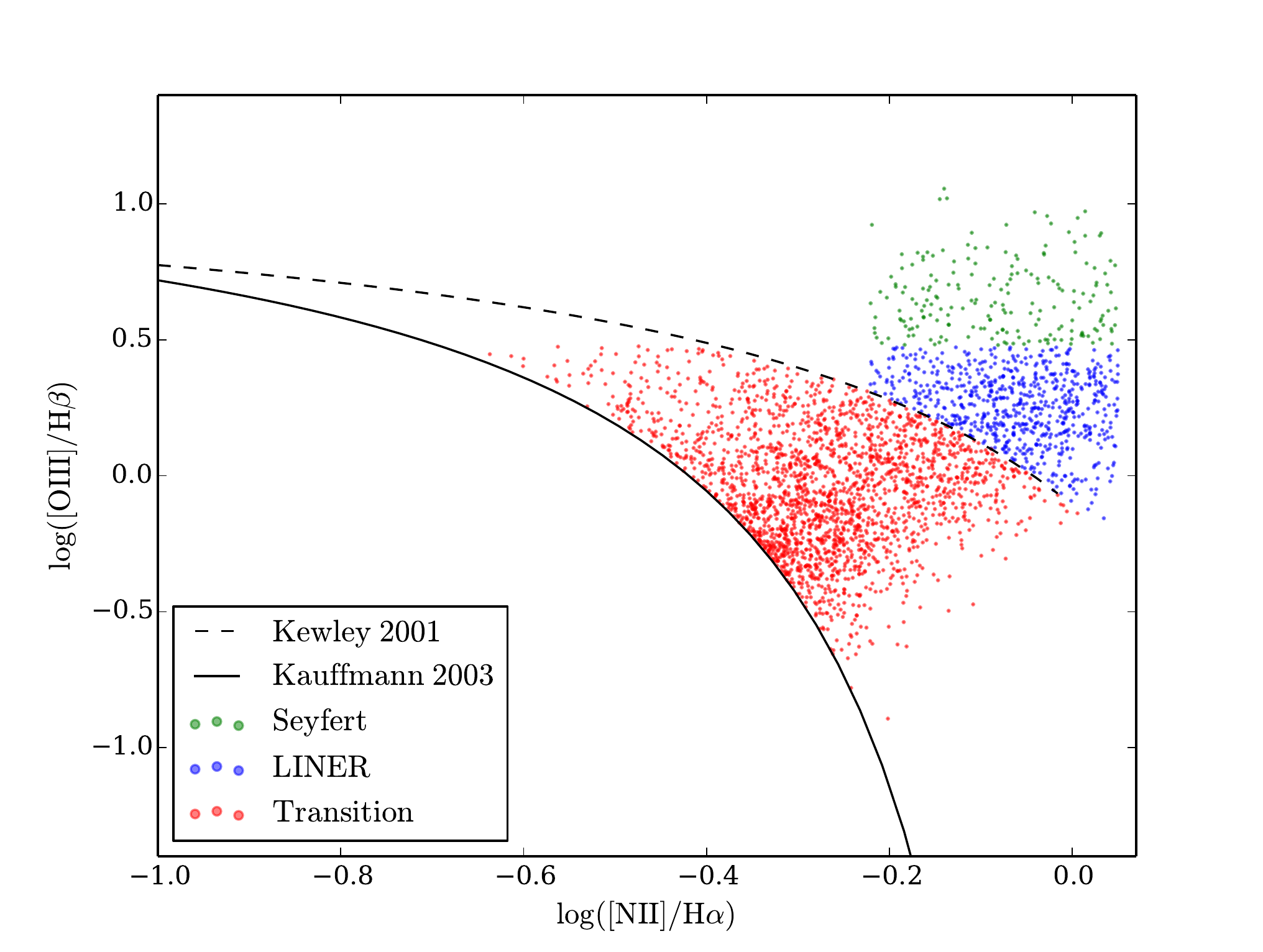}
\caption{BPT diagram to classify AGN into Seyferts, LINERs, and Transition objects.\label{fig:diag}}
\end{figure}

\begin{table}
\centering
\begin{tabular}{ccc}
Spectral type & Definition \\ \hline
Transition object & Starburst galaxy \\
Seyfert &  AGN with [OIII]/H$\beta$ > 3 \\
LINER &  AGN with strong low ionization lines \\
\end{tabular}
\caption{Spectral classification of type 2 AGN. \label{tab:classification}}
\end{table}

    \cite{Bald81} have shown that it is possible to distinguish type 2 AGN from normal star-forming galaxies by the intensity ratios of relatively strong emission lines. The MPA/JHU database contains the intensity ratios [OIII]$\lambda$5007/H$\beta$ and [NII]$\lambda$6583/H$\alpha$ (hereafter [OIII]/H$\beta$ and [NII]/H$\alpha$) that allow us to classify AGN according to the Baldwin, Phillips \& Terlevich (BPT) diagram.

\begin{table}
\centering
\begin{tabular}{ c c c }
Spectral type 		& Total & Weakly accreting \\ \hline
Seyferts 			& 197	& 33 \\ 
LINERs 				& 683	& 322 \\ 
Transition objects	& 2176 	& 1400\\  \hline
\end{tabular}
\caption{Number of objects of each spectral type used in this study. We limit the study to $z<0.041$ (120 h$^{-1}$ Mpc). \label{tab:numbers}}
\end{table}

    The [OIII] line can be excited by AGN as well as by massive stars but it is known to be weak in metal-rich star-forming regions. However, for star-forming galaxies, the [OIII]/H$\beta$ ratio increases while the [NII]/H$\alpha$ ratio increases in high gas-phase metallicities \citep{Char2001}. 
    
    In the BPT diagram, AGN lie above the extreme starburst line defined by \cite{Kewl01} and star-forming galaxies are below the \cite{Kauf03c} line. Objects between the lines host a mixture of star formation and nuclear activity and are classified as Transition objects. The objects where the AGN component is dominant can be split into two groups: those with [OIII]/H$\beta > 3$ are Seyferts and the rest are classified as Low-Ionization Nuclear Emission-line Regions (LINER) \citep{Kauf03c}. \ LINERs typically have lower luminosities than Seyferts and their low ionization lines such as [OI] or [NII] are relatively strong. LINERs spectra could be produced in cooling flows, starburst-driven winds and shock-heated gas \citep{Filippenko92} and this opened a debate whether LINERs should be considered a low-luminosity extension of the AGN sequence \citep{Ho2003}. Table \ref{tab:numbers} shows the number of objects of each spectral type used in this study, up to redshift $z<0.04$ (120 h$^{-1}$ Mpc).
    
   We follow \cite{Kauf03c} classification criteria, as shown in Fig. \ref{fig:diag} and Table \ref{tab:classification}. The advantage of this criteria is that it is less sensitive to the projected aperture size of the fibers than other classification schemes based on lower ionization lines \citep{Kauf03c}. 
   
\section{Results}

	\subsection{AGN abundance} \label{sec:abundance}
    Since AGN can be observed at large distances,  they are believed to be ideal tracers of the structure of the Universe at the largest scales. AGN played a key role in the early phases of cosmological studies but it was shown by source counts that AGN population evolve with cosmic epoch \citep{Merloni2012}. Hence, the cosmological information, such as the geometry or the cosmological parameters, is masked by the evolution of AGN themselves. In this section, we study the abundances of AGN in different density environments and compare those to the corresponding galaxy abundances.

    Figure \ref{fig:abundances} shows the abundances, the number density of objects as a function of the environmental density (Section \ref{ann:abundance}), for the AGN catalog (MPA/JHU) and two galaxy catalogs (2M++ and SDSS) limited in the same redshift range $z<0.04$. In order to compare the results of the different catalogs, we rescaled the curves by dividing by the abundance in the first bin. 
       
    As can be seen, the abundances show the same trend for AGN and galaxies.  This result is compatible with the findings of \cite{Jiang2016} and \cite{Li2006} who studied the cross-correlation function of AGN and a reference galaxy sample and concluded that the AGN clustering signal is the same as that of galaxies at large scales (>1 Mpc h$^{-1}$) indicating that the halos of active and inactive galaxies have similar masses. They found some differences in the number of neighbors of AGN in small scales (<100 kpc h$^{-1}$) but our study is limited to scales of 2.6 Mpc h$^{-1}$ which is the resolution of the reconstructed density field. 
       
\begin{figure}[t]
\centering
\includegraphics[width=3.7in]{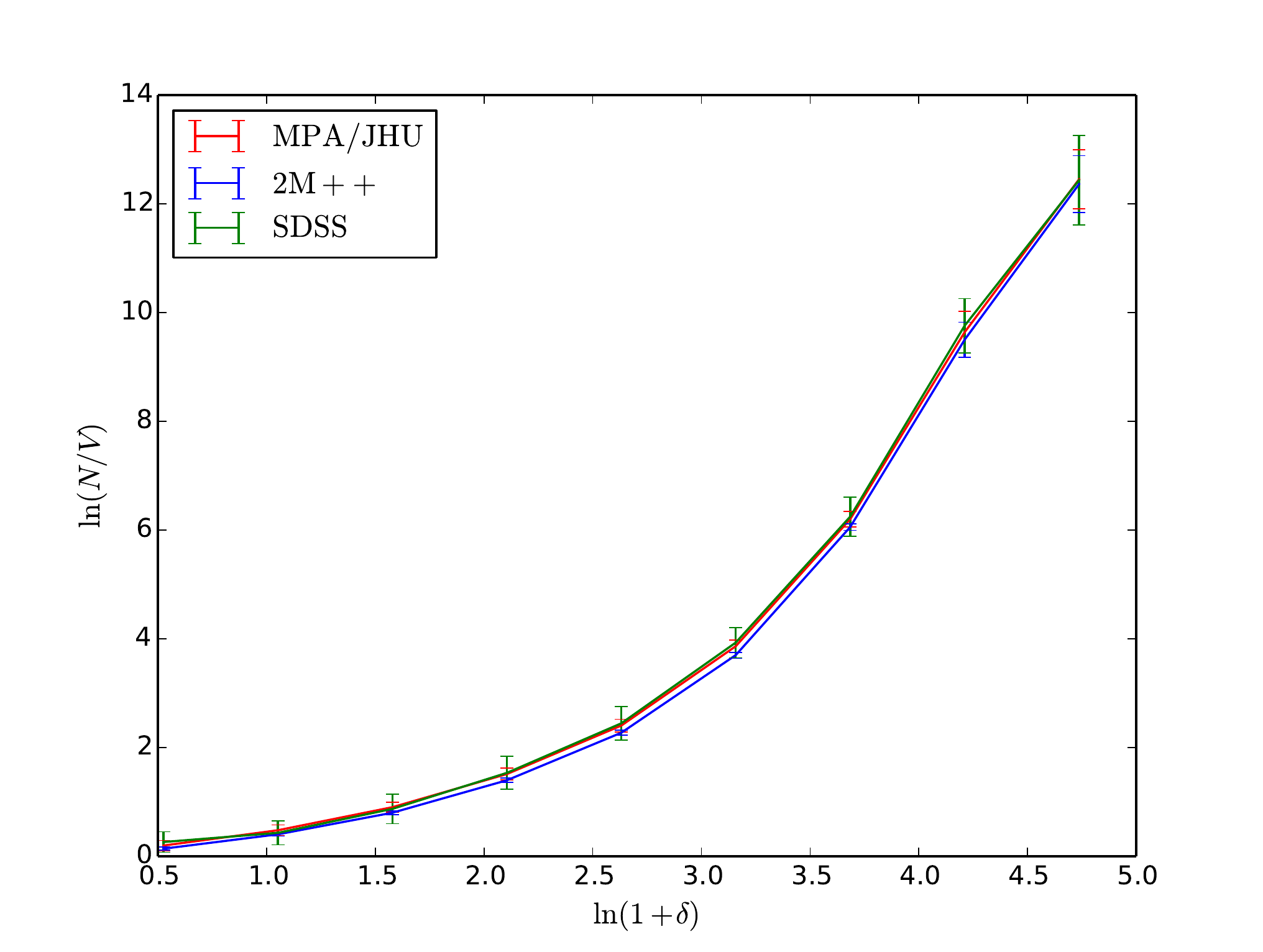}
\caption{Abundance of AGN and galaxies. The curves are rescaled in order to compare the results for different catalogs. Galaxies and AGN show the same abundances which suggests that their halos have similar masses.  \label{fig:abundances}}
\end{figure}

    \subsection{Occurrence rate}
   	The evolutionary sequence suggested by \cite{Const08} was derived from the study of the occurrence rate of different spectral types in different environments (defined in Section \ref{ann:abundance}). Since the dynamical evolution in underdensities is slower, the fraction of different spectral types can provide some information on AGN evolution. 
    
    The upper panel in Fig. \ref{fig:occ} shows that Seyferts and Transition objects have larger occurrence rates in underdensities than in clusters while LINERs are equally represented in under and over densities. This might indicate an evolutionary sequence since  Seyferts and Transition objects in very high density regimes, with faster dynamical evolution, have already evolved into LINERs. This is also consistent with the increase of LINERs occurrence rate in clusters. 
    
\begin{figure}[t]
\centering
\includegraphics[width=3.7in]{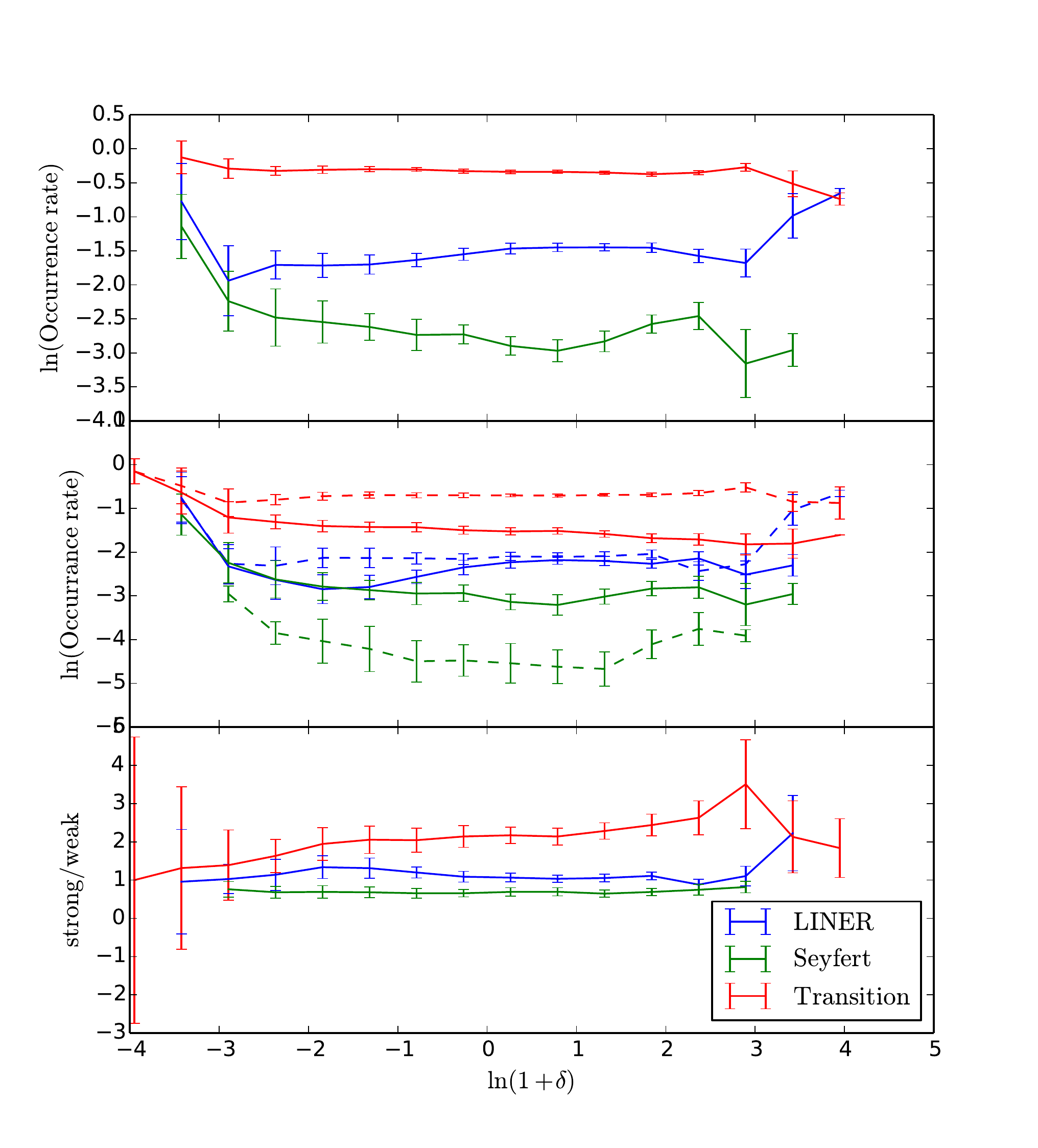}
\caption{Occurrence rate for different spectral types as a function of the density contrast. The upper panel shows that Transition objects are less abundant in high densities while LINERs show opposite trend. The middle and bottom panels show that strongly (solid line) accreting Transition objects (defined as  $L_\mathrm{[OIII]}>10^{39}$ erg s$^{-1}$) are more sensitive to density contrast than weakly accreting (dashed line). This is compatible with the evolutionary sequence suggested by \cite{Const08} with Transition objects transforming into LINERs}. \label{fig:occ}
\end{figure}

	Since the dynamical evolution can be faster for highly accreting objects, we study the effect of the accretion rate on the occurrence rates. Following \cite{Const08}, we consider that objects with $L_\mathrm{[OIII]}>10^{39}$ erg s$^{-1}$ are weakly accreting. The middle panel in Fig. \ref{fig:occ} shows that weakly accreting Transition objects are equally represented in any density regime while their counterparts with high accretion objects become less frequent with density. This might indicate that Transition objects with high accretion rate make their transformation to another spectral type faster. We can see that weakly accreting Seyferts show stronger dependences with density than high accreting ones. LINERs do not show a significant difference between high and low accretion rates except at very high densities, where weakly accreting objects are more abundant. This higher occurrence of weak LINERs at very high densities might be related to their position in the cluster: the highest densities are found in the center of clusters, where the interaction with late-type galaxies containing gas is less likely and hence the gas accretion might be reduced \citep{Chris04,Hwang11}.   
        
    In order to compare our results to \cite{Const08} we study the AGN occurrence rates for voids and walls. Following their approach, we define voids according to a density threshold between wall and void overdensity. Table \ref{tab:occ} shows occurrence rates of objects in voids and walls for different density thresholds. As can be seen, the ordering of the occurrence rate between of AGN in wall and void regions do not depend on the density threshold. However, the  difference between voids and walls could be a statistical fluctuation. For this reason, we include the significance of the results in the Table \ref{tab:occ}. The difference between voids and walls is considered to be significant when
    \begin{equation}
    |\mathrm{Occurrence}_\mathrm{wall}-\mathrm{Occurrence}_\mathrm{void}| > 2\sqrt{\epsilon_\mathrm{void}^2+\epsilon_\mathrm{wall}^2},
    \end{equation}
    being $\epsilon$ the uncertainty in the occurrence rate.
    We can see that the significance of the results strongly depends on the threshold. \cite{Const08} set the threshold at $\delta=-0.6$ and found that Seyferts are equally represented in voids and walls while Transition objects and LINERs show a preference for walls. We can only reproduce their result for Seyferts and LINERs when the threshold is $\delta=-0.7$, showing that the results obtained by density thresholding depend on the details of the definition of the density field.

\begin{table}[tbp]
\centering
\begin{tabular}{cllll}
\begin{tabular}[c]{@{}c@{}}$\delta$ \\ threshold\end{tabular} & \multicolumn{1}{c}{\begin{tabular}[c]{@{}c@{}}Spectral\\ Type\end{tabular}} & Wall            & Void & Significance           \\ \hline
\multirow{3}{*}{-0.7}             & LINER      	& 17 $\pm$ 1  		& 22.2$\pm$0.9   	&	Significant \\
                                  & Seyfert    	& 8$\pm$ 1  		& 6.35 $\pm$ 0.07   &	Not signif.  \\
                                  & Transition 	& 73$\pm$ 2  		& 71.4$\pm$ 0.1		&	Not signif. \\ \hline
\multirow{3}{*}{-0.6}             & LINER  		& 18 $\pm$ 1  		& 22.5 $\pm$ 0.1   	&	Significant \\
								  & Seyfert  	& 8.1$\pm$ 0.7  	& 6.27 $\pm$ 0.09   &	Significant \\
								  & Transition  & 73 $\pm$ 1  		& 71.3$\pm$ 0.2   	&	Not signif. \\ \hline
\multirow{3}{*}{-0.5}             & LINER		& 18$\pm$1  		& 22$\pm$2  		&	Significant \\
								  & Seyfert  	& 7.6 $\pm$ 0.5  	& 6.26 $\pm$ 0.09   &	Significant \\
								  & Transition  & 74$\pm$ 1  		& 71.1 $\pm$ 0.2  	&	Significant \\ \hline
\end{tabular}
\caption{Occurrence rate (in \%) for different density thresholds between void and wall overdensities. We can see that the significance of the results depends on the threshold. \label{tab:occ}}
\end{table}

	\subsection{Cosmic web analysis}
	In the previous section, we studied the occurrence rate as a function of density. In this section we analyze whether the occurrence rate depends on the AGN location in the cosmic web. We classified different environments as voids, sheets, filaments and clusters as described in Section \ref{sec:Ts}.           
    Fig. \ref{fig:ts} shows that Transition objects are relatively more abundant in voids than in clusters while LINERs show opposite trends and Seyferts do not show a clear trend. This might also support that LINERs can be a later stage in the evolutionary sequence and hence they are more abundant where the dynamical evolution is faster due to interaction and merging. 
        
\begin{figure}[t]
\centering
\includegraphics[width=3.7in]{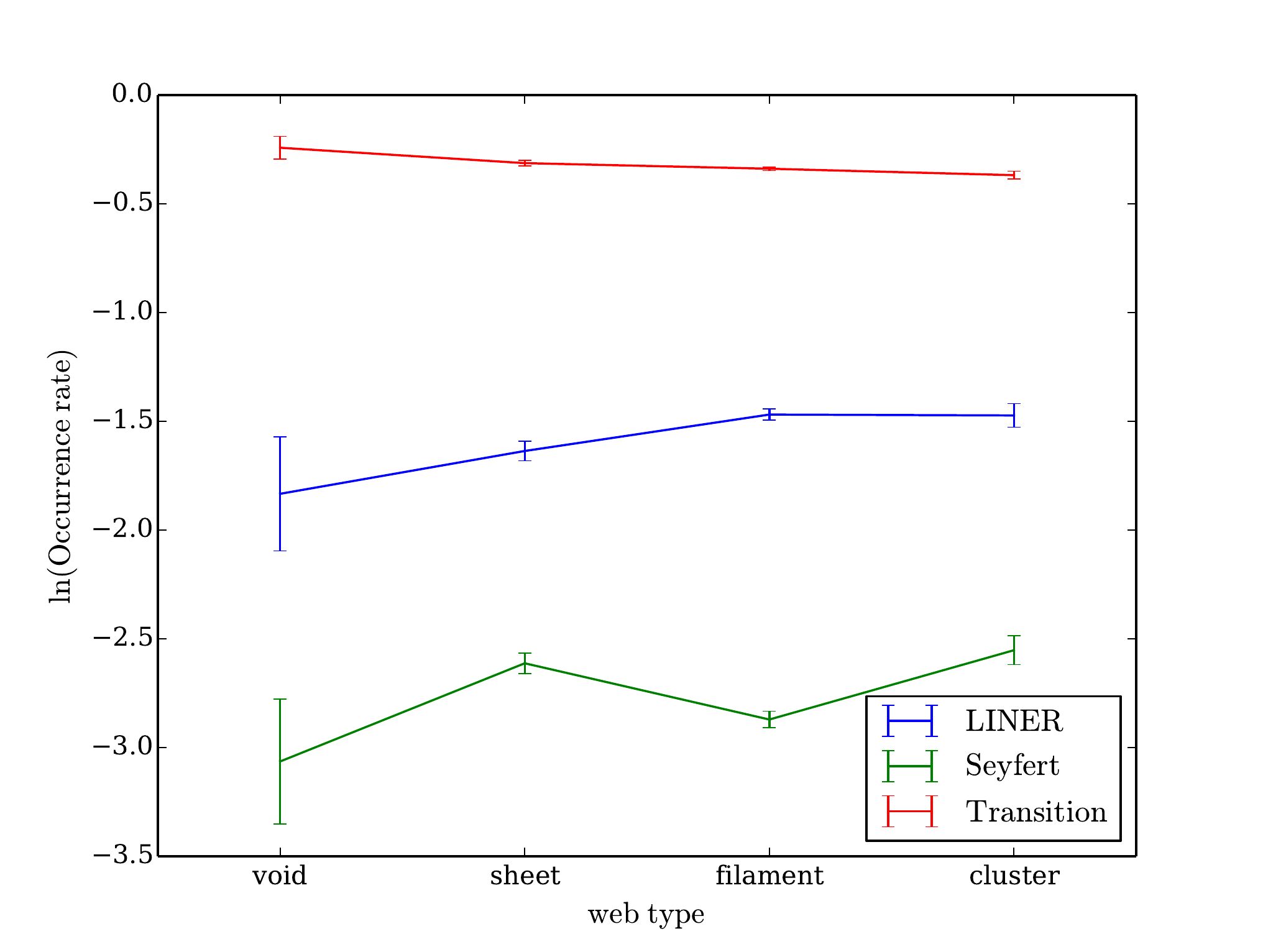}
\caption{AGN ocurrence rate in different structures. Transition objects are more abundant in voids while LINERs are more abundant in clusters which is compatible with Transition objects evolving into LINERs. \label{fig:ts}}
\end{figure}

    \subsection{Properties of AGN and LSS environment}
    We have studied how the AGN formation depends on the environmental density. In this section, we focus on how the large-scale environment affects AGN properties. We studied the dependence of AGN and their galaxy host properties with the environmental density and the web-type structure. While the density is related to the amplitude of the large-scale gravitational potential, web-types provide information on the shape of the large-scale  potential.
    
    \subsubsection{Comparing spectral types}
    The left panels in Fig. \ref{fig:prop} show AGN properties as a function of environmental density for each spectral type. Transition objects show stronger dependency on the density contrast than LINERs and Seyferts. This can be related to their larger amount of gas since the availability of gas at merging can trigger star formation and affect the dynamics of galaxy interaction. 
       
    We can see  in Fig. \ref{fig:prop} that the stellar mass of the host galaxy, $\log(M_*)$, decreases at high densities for Transition objects and Seyferts. A possible explanation for this phenomena is the gas stripping due to interactions with other galaxies. LINERs do not show this behavior because their gas content is smaller. We can also see that galaxy hosts in underdensities are more massive than in the mean density, especially for Transition objects. This indicates that void AGN are hosted in the most massive galaxies in voids. This result is consistent with  \cite{Const08}, where they found that void AGN hosts are not dominated by the abundant less massive galaxies in the underdense regions. 
    
    The surface mass density $\log(\rho_*)$ and the concentration index $C$ allow to study the galaxy morphology \citep{Kauf03c}. These two quantities can be used to separate early-type galaxies (Hubble types Sa, S0, and E) with $C>2.6$ and surface mass density in the range $3\times 10^8 - 3\times 10^9 M_\odot \mathrm{kpc}^{-2}$, from late-type galaxies \citep[spirals and irregulars, ][]{Strateva01}.  Figure \ref{fig:prop} shows that LINERs and Seyferts occupy early-type galaxy hosts in any density regime. Transition objects also prefer early-type hosts in voids. However, the mean of their concentration index is $C\approx 2.6$ for mean and large densities which indicates that the host galaxy of these objects can present different morphologies. We completed the analysis of their morphology in the next section, considering the differences due to accretion rates.
    
    We also study the stellar population of the host galaxy. The 4000\AA{} break is an indicator of the mean age of the stellar population which has smaller values for younger populations. Strong values of H$\delta$ absorption arises due to recent bursts of star formation \citep{Kauf03c}. We can see that Transition object hosts are younger than Seyfert and LINER galaxy hosts. This is consistent with their larger amount of gas and the fact that Transition objects lie between the extreme starburst line and the pure star formation line, meaning that they can still produce new stars. The amplitude of the 4000\AA{} break decreases at highest densities for Transition objects suggesting that strong interaction and merging in clusters trigger star formation in these objects. This is consistent with the H$\delta$ trend since Transition objects show a larger increase of the H$\delta$ absorption at high densities, indicating a more powerful starburst. These two quantities also show that Seyferts in underdensities have a slight preference for younger galaxy hosts while LINER hosts are similar in under and mean density regions but become older in high density regimes, which may be related to the slower dynamical evolution in voids. 
    
    The stellar velocity dispersion $\sigma_v$ is proportional to the black hole mass \citep{Ferr00}. We can see that velocity dispersions are larger in higher densities for LINERs. However, it decreases for high density Transition objects. This could suggest that Transition objects at high densities host less massive black holes because they are still forming. However, Transition objects show a recent starburst in the same density regime that can affect the velocity dispersion. Since new stars are formed in gas clouds, they have the same velocity and this might affect the measurement of stellar velocity dispersions in the galaxy. Hence, velocity dispersion might become a bad estimator for the mass of the central black hole after a recent starburst.

\begin{figure}[t]
\centering
\includegraphics[width=4in]{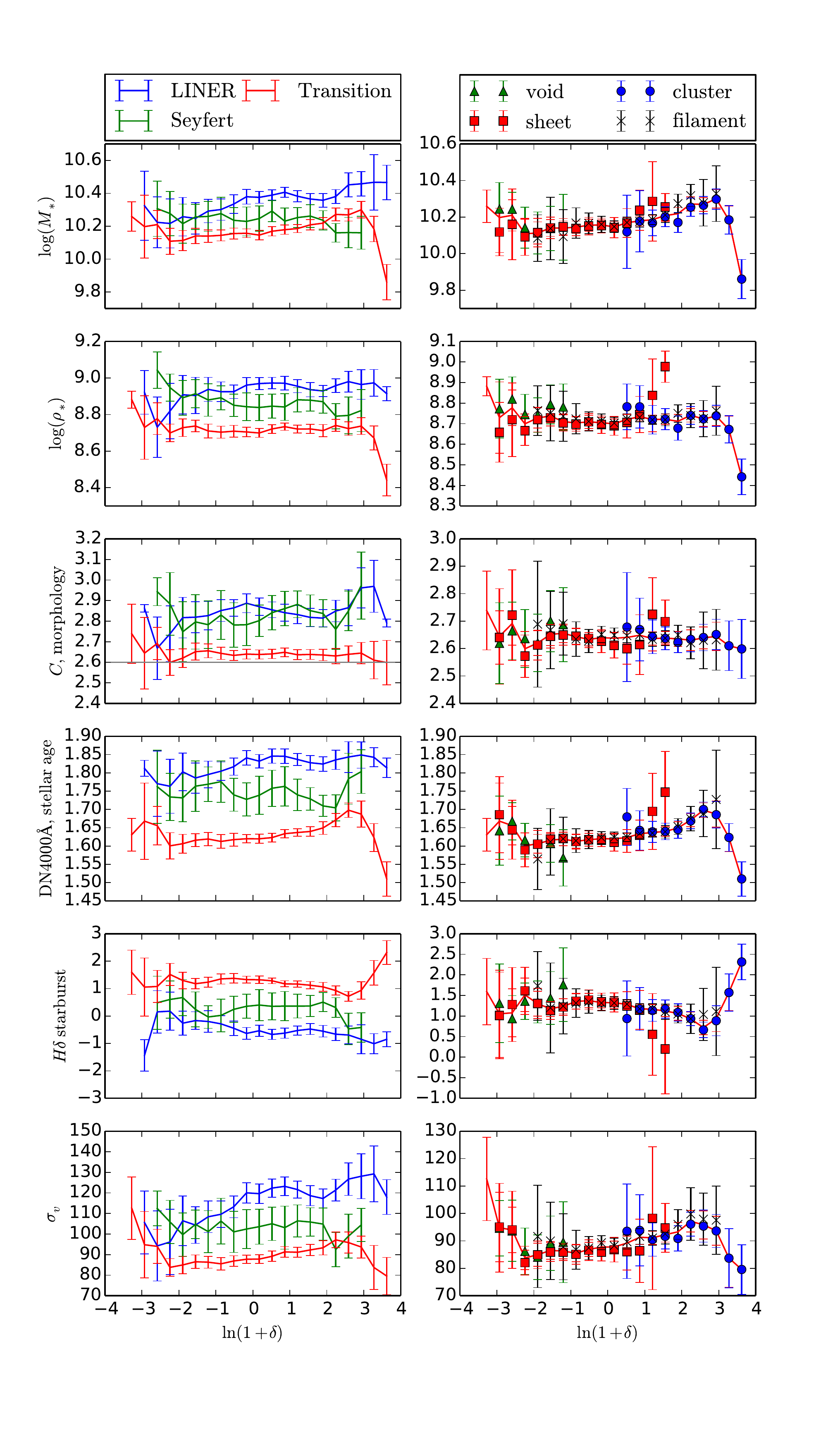}
\caption{Properties of different spectral types of AGN as a function of density contrast. We can see that the stellar mass drops with the density due to gas stripping and high density objects show a stronger starburst and younger stellar populations. The line in $C=2.6$ separates the early- and late-type galaxies, showing that Seyferts and LINERs are found in early-type hosts. The left panels show that AGN properties are more dependent on the density than on the web-type.\label{fig:prop}}
\end{figure}

In order to determine if AGN properties depend on the cosmic web or the density, the right panels in Fig. \ref{fig:prop} show the properties at different web-type structures as a function of density. We can see that AGN properties do not depend on the web-type classification but on the density since different structures show the same values in the same density range. Since these properties are related to the star formation, it is expected that they depend on the density more than on the shape of the gravitational potential. 

	\subsubsection{Comparing accretion rates}
    
Since some authors \citep{Const08,Kauf03c} found that AGN properties depend on the accretion rate, we compared how the environment affects AGN properties in each spectral type for different accretion rates.

Figure \ref{fig:L_LINER} shows that strongly accreting Seyferts have less massive hosts at high densities while Transition objects and LINERs do not show this trend. All spectral types have more massive hosts for high accretion rates. However, Transition objects have massive hosts in voids for any accretion rate.

\begin{figure}[t]
\centering
\includegraphics[width=4in]{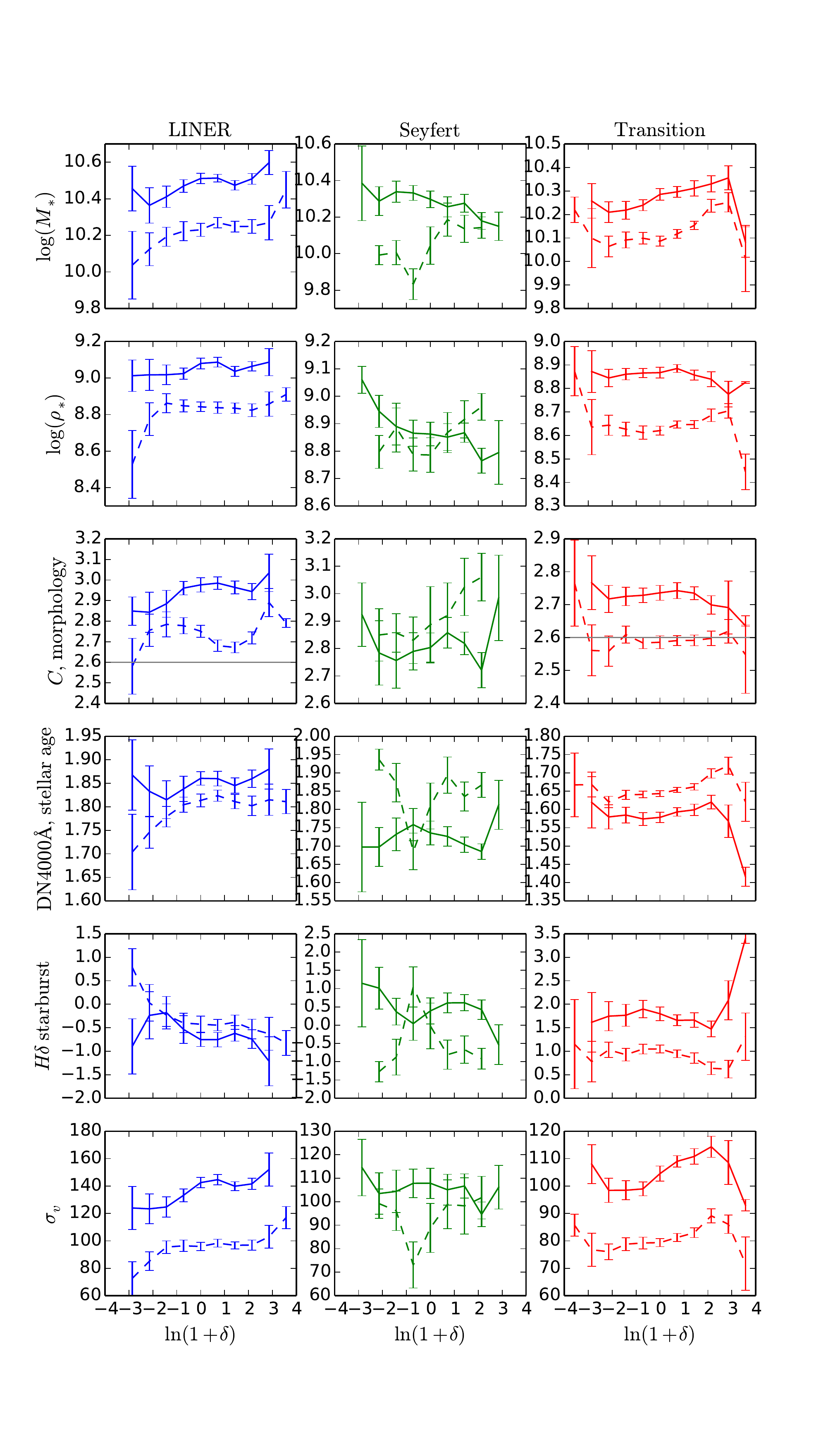}
\caption{AGN properties for each spectral type and reported accretion rates. As can be seen, strongly (solid line) and weakly (dashed) accreting LINERs show a larger difference in underdensities. Transition objects with high accretion rate have more powerful starburst and smaller concentration indices at high densities (irregular galaxies). Seyferts show a transition at $\ln(1+\delta)\approx-0.7$. \label{fig:L_LINER}}
\end{figure}

In the previous section, we found that Seyferts and LINERs prefer early-type galaxies in any density regime while Transition objects can be found in different morphologies. In this section we compare their concentration index and stellar mass density for different accretion rate. Figure \ref{fig:L_LINER} shows that Transition objects with low accretion rate can be found in late type galaxies ($C<2.6$) but they prefer early-type galaxies in underdensities. This can be due to a higher abundance of early-type galaxies in voids. We can also see that the concentration index decreases at high density for objects with high accretion rate. A possible explanation is that the interaction and merging in clusters lead to an irregular galaxy. 

\begin{figure}[t]
\centering
\includegraphics[width=4in]{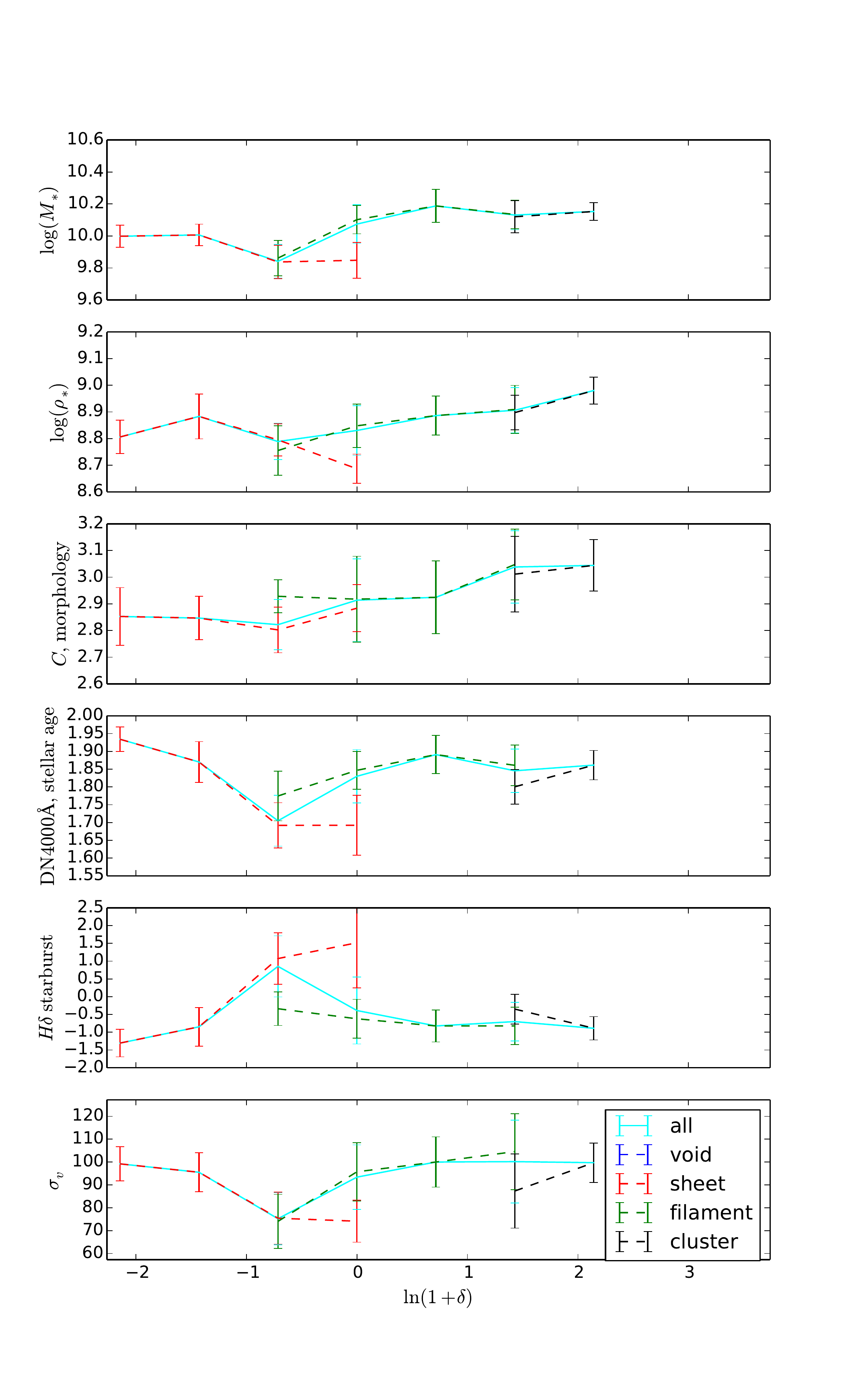}
\caption{Properties of weakly accreting Seyferts. Around $\ln(1+\delta)\approx-0.7$ these exhibit a different behavior if located in sheets or in filaments. This difference in AGN formation times in these two environments might indicate a transition. \label{fig:sy_weak}}
\end{figure}

Transition objects with higher accretion rates show stronger starbursts ($H\delta$) than weakly accreting objects. This is consistent with the 4000\AA{} break, indicating a younger population for Transition objects with a high accretion rate, especially at high densities. LINERs show different behavior only in low density regime while at high densities the starburst and stellar age is equivalent to high and low accretion rate. Weakly accreting LINERs in voids show a young stellar population and stronger starburst. This might correlate with the concentration index: weakly accreting LINERs in voids are found in late-type galaxies, probably spiral galaxies with a larger amount of gas that can form new stars.

Weakly accreting Seyferts show a transition at $\ln(1+\delta)=-0.7$ in all their properties. In order to study this feature, Fig.~\ref{fig:sy_weak} shows the properties for those objects in different web-type structures. Unlike the other spectral types, weakly accreting Seyferts show a different behavior if located in sheets or in filaments of the same density. This may indicate different populations of Seyferts as a function of web-types of the same density. Specifically objects in sheets are on average younger than in filaments. This may be a signature of recent AGN formation but further investigation with next generation data and more detailed density reconstructions are required to confirm this feature in the future.

\section{Conclusions}
	In this work, we have studied the effect of the large-scale environment on AGN: how the environment affects the formation and properties of AGN. We have used a 3D high-resolution (2.6 Mpc h$^{-1}$) density field obtained from a Bayesian reconstruction applied to the 2M++ galaxy catalog. The study is based on the MPA/JHU AGN catalog \citep{Kauf03c}, which contains only Type 2 AGN. We limit our study to objects within 120 Mpc h$^{-1}$ (z<0.04).
    
    We confirm that the environment affects the formation and properties of AGN. Particularly, AGN properties and formation depend on the environmental density more than on the web-type. Hence, the amplitude of the  large-scale gravitational potential affects AGN more than the shape of the  large-scale  potential. However, Seyferts with low accretion rate show some sensitivity to whether they reside within sheets or filaments.
    
    The AGN abundance is the same as that of galaxies, indicating that halos are similar for active and inactive galaxies. When comparing spectral types and accretion rates we have found differences in occurrence rates. Weakly accreting LINERs are more abundant in underdensities than in clusters while Transition objects show opposite trends. This is consistent with the AGN evolution of Seyferts/Transition objects into LINERs suggested by \cite{Const08}.
    
   AGN properties are also affected by the environmental density. The effect of the environment is stronger for Transition objects. This might be related to their larger amount of gas. It was found that the stellar mass of the host grows with the environmental density only for LINERs while Seyferts and Transition objects show a decrease of their stellar mass at very high densities. This might be a result of the gas stripping due to galaxy interaction and merging. It is also interesting that the AGN population in voids is not dominated by low-mass hosts but AGN are found in the most massive void galaxies. The starburst and age of the stellar population are also affected by the large-scale environment: younger populations and more powerful starbursts are found in clusters. This might be an indicator of the interactions and merging that trigger the star formation in AGN hosts. 
    
    We also find that the effect of the environment on AGN properties is different for weakly accreting objects. For instance, the morphology of Transition objects depend on their accretion rate, showing that weakly accreting objects are found in late-type galaxies. Seyferts with low accretion rate are younger in sheets than those in filaments in the same density regime, which might indicate different populations in these two environments.
    
    These results indicate some particular properties of void AGN to be confirmed and studied by larger AGN samples of next generation surveys.
      
\section*{Acknowledgements}
This research was supported by the DFG cluster of excellence "Origin and Structure of the Universe" (www.universe-cluster.de). This work was supported in part by the ANR grant ANR-16-CE23-0002.

\appendix

\section{Aligning observed AGN with inferred density fields} \label{ann:align}
In order to study the large-scale environment of observed AGN, we needed to relate them to our reconstructions of the three-dimensional density field. This means it was necessary to map their coordinates into our analysis domain. 

Since the three-dimensional density fields have been inferred with respect to the CMB restframe, we first had to adjust redshifts of observed objects correspondingly in the MPA/JHU catalog. The transformation is achieved by accounting for the observers velocity with respect to the CMB $v_\mathrm{los}$ along the line of sight $v_\mathrm{los}$:
\begin{equation} 
z = z_\mathrm{obs} + v^\mathrm{CMB}_\mathrm{los}/c, 
\end{equation}
where the line of sight velocity is given by \cite{Tull2007}:
\begin{equation}
v^\mathrm{CMB}_\mathrm{los} = -25\cos l\cos b -246\sin l \cos b +277\sin b,
\end{equation}
with $(l,b)$ being the galactic coordinates and $v_\mathrm{los}$ is obtained from the projection of $v_\mathrm{CMB} = (-25,-246,277)$ km s$^{-1}$  which is the relative velocity of the sun with respect to the CMB in Galactic coordinates. We then translated corrected redshifts into comoving distances $d$ by solving the equation: 
\begin{equation}
d = d_H \int^z_0  \frac{dz'}{E(z')},
\end{equation} 
where we have used the cosmological parameters given in Section \ref{sec:method}.
Finally we transformed spherical to Cartesian coordinates via:
\begin{equation}
\vec{r} = d(\cos l \cos b, \cos b \sin l, \sin b).
\end{equation}
Corresponding grid positions of the analysis domain are then obtained by rescaling the coordinate vector $\vec{r}$ to the voxel index as $\vec{r}_\mathrm{pix} = (N/L)\vec{r}$ where $N$ is the number of voxels in each dimension and $L$ is the size of the reconstruction domain. Since the observer is at the center of the box, $\vec{r}$ is shifted by $L/2$.


\bibliographystyle{aa} 
\bibliography{biblio} 
\end{document}